\documentclass[12pt]{article}
\usepackage{epsfig}
\newcommand{\simlt}  {\raisebox{-.6ex}{$\stackrel{\textstyle <}{\sim}$}}

\newcommand{\Emt}{E_{\mu\tau}}
\newcommand{\mtrs}{$\mu-\tau$ reflection symmetry}
\newcommand{\numu}{\nu_{\mu}}
\newcommand{\numubar}{\overline{\nu}_{\mu}}
\newcommand{\nutau}{\nu_{\tau}}
\newcommand{\nutaubar}{\overline{\nu}_{\tau}}
\newcommand{\uu}{\underline{u}}
\newcommand{\uv}{\underline{v}}
\def\ni{\noindent}
\def\nl{\hfill\break}
\begin{document}
\begin{flushright}
hep-ph/0210197 \\
RAL-TR-2002-023 \\
30 Sep 2002 \\
\end{flushright}
\begin{center}
{\Large $\mu-\tau$ Reflection Symmetry in Lepton Mixing and}
\end{center}
\vspace{-7mm}
\begin{center}
{\Large Neutrino Oscillations}
\end{center}
\vspace{1mm}
\begin{center}
{P. F. Harrison\\
Physics Department, Queen Mary University of London\\
Mile End Rd. London E1 4NS. UK \footnotemark[1]}
\end{center}
\begin{center}
{and}
\end{center}
\begin{center}
{W. G. Scott\\
Rutherford Appleton Laboratory\\
Chilton, Didcot, Oxon OX11 0QX. UK \footnotemark[2]}
\end{center}
\vspace{1mm}
\begin{abstract}
\baselineskip 0.6cm
\noindent
We examine the possibility that the lepton mixing matrix may exhibit
a strong form of $\mu-\tau$ universality, in which corresponding elements
of the $\mu$- and $\tau$-neutrino flavour eigenstates have equal moduli,
and find that it is consistent with present data on neutrino oscillations.
We point out that in the Dirac case, this is equivalent to symmetry under 
$\mu-\tau$ reflection, ie.~the combined operation of $\mu-\tau$ flavour 
exchange in the MNS matrix with a $CP$ transformation on the whole leptonic 
sector. 
We give the most general form for such a mixing matrix, examine the lepton
mass matrices under such a symmetry, and explore the observable manifestations 
of the symmetry in neutrino oscillations.
\end{abstract}

\begin{center}
({\it To be published in Physics Letters B})
\end{center}

\footnotetext[1]{E-mail:p.f.harrison@qmul.ac.uk}
\footnotetext[2]{E-mail:w.g.scott@rl.ac.uk}

\newpage
\ni {\bf 1 Strong $\mu-\tau$ Universality in Lepton Mixing}
\vspace{2mm}
\nl The data on atmospheric neutrinos \cite{atmos} \cite{superKElec} indicate 
that the muon neutrino couples to the heaviest neutrino mass eigenstate, 
$\nu_3$, with a strength $|U_{\mu3}|\simeq 1/\sqrt{2}$, where $U$ is the 
conventional MNS lepton mixing matrix \cite{mns}. Independently, the data 
from the CHOOZ \cite{chooz} and PALO VERDE \cite{paloverde} reactor neutrino 
experiments indicate that the electron neutrino has at most a 
small coupling to the heaviest neutrino, $|U_{e3}|^2~\simlt~0.03$. 
These two facts taken together with the normalisation condition on the $\nu_3$
eigenstate imply that $|U_{\tau3}|\simeq 1/\sqrt{2}$ and hence, in particular, 
that $|U_{\mu3}|\simeq|U_{\tau3}|$. 
Since $U_{e3}$ is small and $|U_{\mu3}|\simeq|U_{\tau3}|$, 
it follows immediately from the orthogonality of the $\nu_3$ eigenstate 
with the $\nu_1$ and $\nu_2$ eigenstates respectively that 
$|U_{\mu i}|\simeq|U_{\tau i}|$ for $i=1,2$ also.
Thus each of the moduli of the three $\mu$-flavour mixing elements is 
at least approximately equal to that of the corresponding 
$\tau$-flavour mixing element.

In this paper, we examine the possibility that
\begin{equation}
|U_{\mu i}|=|U_{\tau i}|,~{\rm exactly,~for~all}~i=1,2,3
\label{equality}
\end{equation}
and find that it is consistent with all the present data on neutrino 
oscillations (excluding the LSND results \cite{lsnd}, which are awaiting 
independent confirmation).
This strong form of $\mu-\tau$ universality (as distinct from the familiar 
lepton flavour universality implicit in the Standard Model, in which
$|U_{\alpha 1}|^2+|U_{\alpha 2}|^2+|U_{\alpha 3}|^2=1,~\forall~\alpha$) 
is interesting, as it removes two of the otherwise 
free parameters of the lepton mixing matrix. 
Several previously suggested hypotheses for the lepton mixing matrix satisfy
Eq.~(\ref{equality}), in particular, the 
trimaximal \cite{tri, HPS33}, 
bimaximal \cite{obm} 
and 
Altarelli-Ferruglio \cite{af} models,
(as well as the more recently proposed
tri-bimaximal \cite{tbm} and
tri-$\chi$maximal \cite{trichim} schemes).
The generalisation of
these hypotheses to a unitary mixing matrix simply satisfying 
Eq.~(\ref{equality}) was already identified as interesting in 
\cite{trichim} and such a mixing matrix was implicit in \cite{ahluwalia}. 
In this paper, we examine this class of mixing matrices and the 
associated phenomenology in more detail.

\vspace{7mm}
\ni \boldmath{\bf 2 Mixing Matrix Parameterisation and $\mu-\tau$ Reflection Symmetry}\unboldmath
\vspace{2mm}
\nl We develop here a simple parameterisation of mixing matrices satisfying 
Eq.~(\ref{equality}) in terms of only moduli of the mixing matrix elements, 
as distinct from the standard parameterisation in terms of mixing angles 
and a phase \cite{pdg}. A unitary matrix cannot contain two identical 
rows. Hence, in a unitary mixing matrix satisfying Eq.~(\ref{equality}), 
the elements of the two rows corresponding to the $\numu$ and $\nutau$ 
respectively must contain relative phases. For Dirac neutrinos, we are free 
to rephase independently the rows and columns of the mixing matrix with
no observable consequence, and we can exploit this freedom (for columns) 
to arrange that $U_{\mu i}=U_{\tau i}^*$ for each such pair of elements. 
We can then write:
\begin{eqnarray}
U=
\left(\matrix{ \underline{u}   \hspace{0.2cm} \cr
               \underline{v}   \hspace{0.2cm} \cr
               \underline{v}^* \hspace{0.2cm} } \right)
=\left( \matrix{ u_1  &
                      u_2 &
                              u_3 \cr
                 v_1 &
                      v_2 &
                              v_3  \cr
      \hspace{2mm} v_1^* \hspace{2mm} &
         \hspace{2mm}  v_2^* \hspace{2mm} &
           \hspace{2mm} v_3^*  \hspace{2mm} \cr } \right)
\label{matrix}
\end{eqnarray}
which is sufficient to ensure Eq.~(\ref{equality}).
The normalisation of any column of $U$ leads to
\begin{equation}
|v_{i}|=\left(\frac{1-|u_{i}|^2}{2}\right)^{1/2},\quad\forall~i
\label{normalisation}
\end{equation}
while the orthogonality of any pair of columns gives
\begin{equation}
u_{i}u_{j}^*=-2\,{\rm Re}(v_{i}v_{j}^*)\quad\forall~i\neq j.
\label{orthogonality}
\end{equation}
Eq.~(\ref{orthogonality}) can be satisfied only if all three elements of 
$\underline{u}$ have a common phase and we use the freedom to rephase 
the electron neutrino flavour eigenstate to set $\underline{u}$ real 
and positive. Using Eq.~(\ref{normalisation}), we may now re-write 
Eq.~(\ref{orthogonality}) as
\begin{equation}
\cos{\alpha_{ij}}=\frac{-u_{i}u_{j}}{(1-u_{i}^2)^{1/2}(1-u_{j}^2)^{1/2}}\quad\forall~i\neq j
\label{orthogonality2}
\end{equation}
where $\alpha_{ij}$ is the relative phase between $v_{i}$ and 
$v_{j}$\footnote{Unitarity triangles formed from the orthogonality
condition of $\uu$ with $\uv$ or $\uv^*$ are always acute-angled triangles.
Those formed from pairs of columns are isosceles triangles.}.
The row $\underline{v}$ (and hence also the row $\underline{v}^*$)
is now completely determined up to an overall phase
by the magnitudes of the elements of $\underline{u}$ (only two of which 
are independent, $\underline{u}$ being normalised to unity).

If the MNS matrix is a matrix of the form, Eq.~(\ref{matrix}), with 
$\underline{u}$ real, the standard action is invariant under the simultaneous 
interchange of the $\numu$ and $\nutau$ flavour eigenstates 
($\uv$ and $\uv^*$ respectively) and complex 
conjugation of the MNS matrix. However, we know 
\cite{cecilia} that complex conjugation of the MNS matrix is formally
equivalent to making a $CP$ transformation on the whole leptonic
sector of the Standard Model (modified to include Dirac neutrino masses and 
standard MNS mixings). We will refer to the combined operation of $\mu-\tau$ 
flavour exchange in the MNS matrix and $CP$ transformation on the leptonic 
sector as ``{\em $\mu-\tau$ reflection}''. Symmetry under $\mu-\tau$ reflection 
implies an MNS matrix which may be written with the $\nu_e$ flavour
eigenstate real and the $\nu_{\mu}$ and $\nu_{\tau}$ flavour eigenstates 
complex conjugates of each other, Eq.~(\ref{matrix}). 
This is sufficient to ensure the strong $\mu-\tau$ universality expressed 
by Eq.~(\ref{equality}), the two definitions being equivalent (for Dirac 
neutrinos\footnote{For Majorana neutrinos, $\mu-\tau$ reflection symmetry 
is more 
constraining on the observable parameters of the mixing matrix than the 
strong $\mu-\tau$  universality of Eq.~(\ref{equality}). In the
following sections of this paper however, there 
is no distinction between the Majorana and Dirac cases, and the 
universality and the symmetry may be taken to be equivalent.}).
As we show at the end of this section, $\mu-\tau$ reflection symmetry is 
consistent with all confirmed neutrino oscillation measurements currently 
available.

For a mixing matrix which is $\mu-\tau$ reflection symmetric, as in
Eq.~(\ref{matrix}), Jarlskog's $CP$-violating invariant \cite{jarls} 
is given by 
\begin{eqnarray}
{\cal J}&=&{\rm Im}(u_{i}u_{j}^*v_{i}^*v_{j})\cr
&=&\frac{1}{2}u_{i}u_{j}(1-u_{i}^2)^{1/2}(1-u_{j}^2)^{1/2}\sin{\alpha_{ij}}\cr
&=&\frac{1}{2}u_{i}u_{j}(1-u_{i}^2-u_{j}^2)^{1/2}\cr
&=&\frac{1}{2}u_1u_2u_3,
\label{JCP}
\end{eqnarray}
a strikingly simple result (we recall the fact that the $u_i$ have been
taken to be real).
We note that quite generally, 
for a given, fixed pair of magnitudes of elements of the electron flavour 
eigenstate, $\underline{u}$, {\em requiring $\mu-\tau$ reflection symmetry, 
is equivalent to requiring maximal $CP$ violation in the leptonic sector}
(the proof is not given here but follows readily by differentiation 
of the complete expression for ${\cal J}$ keeping $|u_2|$ and $|u_3|$ fixed; 
it may also been seen in the standard parameterisation \cite{pdg}, 
where Eq.~(\ref{equality}) corresponds to fixing $\theta_{23}=\pi/4$ and 
$\delta=\pm \pi/2$).

In order to make mixing matrices of the form Eq.~(\ref{equality}) 
$\mu-\tau$ reflection symmetric, we have defined the phase convention 
of the first row and all the columns, and maintaining these conventions 
restricts somewhat our freedom to rephase the remaining rows of the matrix. 
However, we are still free to make a phase transformation on the matrix
$U\rightarrow \Phi U$ where $\Phi=\exp\{i~{\rm diag}(0,\phi,-\phi)\}$, 
without any observable consequence, and without disturbing the
$\mu-\tau$ reflection symmetry.
We may therefore choose $\phi$ such that any chosen neutrino mass eigenstate 
(any column) is wholly real. With such a phase convention, the matrix 
$U$ has only two free parameters, which can be taken to be any pair 
of the (real) elements of $\underline{u}$.

It is perhaps of the most practical convenience to choose $\phi$ such that 
the $\nu_3$ eigenstate is real (so that the three parameters presently
constrained directly by experiment, $U_{e2}$, $U_{e3}$ and $U_{\mu3}$
are all real) and to take as the two free parameters, 
$U_{e2}=u_2$ and $U_{e3}=u_3$. We know from the HOMESTAKE \cite{homestake},
SUPER-K \cite{sksolar} and SNO \cite{sno} solar neutrino 
meaurements that $|U_{e2}|\simeq 1/\sqrt{3}$ and from the 
CHOOZ \cite{chooz} and PALO VERDE \cite{paloverde} reactor experiments 
that $|U_{e3}|~\simlt~0.17$. Hence, the hypothesis of $\mu-\tau$ reflection 
symmetry accommodates all atmospheric, solar, and reactor neutrino data 
currently available (each of these results is confirmed by more than 
one experiment). The LSND result \cite{lsnd}, cannot be accommodated
simultaneously with all these confirmed data using a $3\times 3$ MNS 
lepton mixing matrix, and would require a fourth family of 
neutrinos \cite{fourthfamily}. We have not considered the question of \mtrs\ 
in such an extended picture.

\vspace{7mm}
\ni \boldmath{\bf 3 $\mu-\tau$ Reflection Symmetry in the Flavour Basis}\unboldmath
\vspace{2mm}
\nl Any symmetry of the lepton mixing matrix should have its origin in 
a corresponding symmetry of the lepton mass matrices expressed in a weak 
basis, namely a basis in which the weak interaction is flavour diagonal
and universal. Such a basis is not uniquely defined however, and we choose 
to consider the special case\footnote{Another special case, the ``circulant 
basis'', was already considered in \cite{trichim}, and a third alternative, 
in which the neutrino mass matrix is diagonal, is examined in Appendix A.} 
in which the charged lepton mass matrix is diagonal (the flavour basis), 
which is defined up to some unobservable phase transformation. We will 
furthermore consider the Hermitian squares, $M^2\equiv MM^{\dag}$, of the 
mass matrices, $M$, in this basis, in order to avoid an arbitrariness 
in the determination of the mass matrices themselves due to the 
non-observation of right-handed charged currents.

In the case of \mtrs\ (ie.~assuming $U$ has the form Eq.~(\ref{matrix}), 
with $\underline{u}$ real), the neutrino mass matrix (squared) is given in the 
flavour basis by
\begin{equation}
M^2_{\nu}= UD^2_{\nu}U^{\dag}
\label{nudef}
\end{equation}
where $D^2_{\nu}={\rm diag}(m^2_1,m^2_2,m^2_3)$, is the diagonal 
neutrino mass matrix squared. Then
\begin{eqnarray}
M^2_{\nu}= \left( \matrix{ z  &  w & w^* \cr
                        w^* &  x &  y  \cr
      \hspace{2mm} w \hspace{2mm} &
         \hspace{2mm}  y^* \hspace{2mm} &
           \hspace{2mm} x  \hspace{2mm} \cr } \right),
\label{numass}
\end{eqnarray}
a form \cite{lavoura} which again explicitly respects $\mu-\tau$ 
reflection symmetry (in this basis, the rows and columns of $M^2_{\nu}$
are labelled by the flavour indices $(e,\mu,\tau)$):
\begin{equation}
M^2_{\nu}\rightarrow (M^2_{\nu})'\equiv \Emt M^2_{\nu}\Emt^{\dag}=(M^2_{\nu})^*, 
\label{mtexchange}
\end{equation}
where
$\Emt$ is the $\mu-\tau$ flavour exchange operator:
\begin{eqnarray}
\Emt= \left( \matrix{ 1  &  0 & 0 \cr
                   0  &  0 & 1 \cr
      \hspace{2mm} 0 \hspace{2mm} &
         \hspace{2mm}  1 \hspace{2mm} &
           \hspace{2mm} 0  \hspace{2mm} \cr } \right).
\label{permutation}
\end{eqnarray}

$M^2_{\nu}$ has 5 observable degrees of freedom: 
$z, x, |w|, |y|$ and the phase of the combination $w^2y$, which together 
correspond to the three neutrino masses, and the two mixing degrees of freedom 
in the matrix $U$. In general, $CP$ violation enters when the combination 
$w^2y$ has a non-trivial phase: 
${\rm Im}(w^2y)=\Delta m^2_{21}\Delta m^2_{31}\Delta m^2_{32}{\cal J}$
(where $\Delta m^2_{ij}=m^2_i-m^2_j$).
The elements of $M^2_{\nu}$ are given by
\begin{eqnarray}
z&=&m^2_1+\Delta m^2_{21}u_2^2+\Delta m^2_{31}u_3^2\cr
x&=&m^2_1+\Delta m^2_{21}|v_2|^2+\Delta m^2_{31}|v_3|^2\cr
w&=&\Delta m^2_{21}u_2v_2^*+\Delta m^2_{31}u_3v_3^*\cr
y&=&\Delta m^2_{21}v_2^2+\Delta m^2_{31}v_3^2.
\label{wxyz}
\end{eqnarray}
These equations may be combined with 
Eqs.~(\ref{normalisation}) and (\ref{orthogonality2}) to 
find closed expressions in terms of any desired pair of elements of 
the vector $\underline{u}$. We have left the unphysical phase of 
$\underline{v}$ (and $\uv^*$) undefined, hence the overall phases of 
$w$ and $y$ are similarly unphysical and undefined, and may be fixed by the 
choice of the phase $\phi$ discussed in Section 2.

Tri-$\chi$maximal mixing \cite{trichim} is a special case of \mtrs\ defined by 
$u_2=1/\sqrt{3}$, $u_3=\sqrt{\frac{2}{3}}\sin{\chi}$, and 
(in a particular phase convention) corresponds to
\begin{eqnarray}
z&=&m^2_1+(1/3)\Delta m^2_{21}+(2/3)\Delta m^2_{31}\sin^2{\chi}\cr
x&=&m^2_1+(1/3)\Delta m^2_{21}+\Delta m^2_{31}[(1/2)-(1/3)\sin^2{\chi}]\cr
w&=&(1/3)\Delta m^2_{21}+\Delta m^2_{31}[-(1/3)\sin^2{\chi}-3i{\cal J}]\cr
y&=&(1/3)\Delta m^2_{21}+\Delta m^2_{31}[-(1/2)\cos^2{\chi}+(1/6)\sin^2{\chi}-3i{\cal J}]
\label{trichi}
\end{eqnarray}
where ${\cal J}=\sin2\chi/(6\sqrt{3})$. 
Tri-bimaximal mixing \cite{tbm} is obtained by setting $\chi=0$.

\vspace{7mm}
\ni \boldmath{\bf 4 $\mu-\tau$ Reflection Symmetry in Neutrino Oscillations}\unboldmath
\vspace{2mm}
\nl The amplitude $A_{\alpha\beta}$ for a neutrino of flavour $\alpha$
to be detected as a neutrino of flavour $\beta$
is given as a function of propagation distance $L$
by the (matrix) equation:
\begin{equation}
A = \exp(-iHL)
\label{expHam}
\end{equation}
where $H$ is the effective neutrino Hamiltonian in the flavour basis
(in the rest of this paper, all calculations are assumed to be in this 
flavour basis). We may take:
\begin{equation}
H=M^2_{\nu}/2E
\label{hamiltonian}
\end{equation}
where $M^2_{\nu}$ is the Hermitian square of the neutrino mass matrix 
with $\mu-\tau$ reflection symmetry given in Eq.~(\ref{numass}), 
and $E$ is the neutrino energy. Obviously, 
since $M^2_{\nu}$ is $\mu-\tau$ reflection symmetric, 
then so is $H$, and we note further that any power or polynomial 
(with real coefficients) of $H$ shares this symmetry.

Eq.~(\ref{expHam}) may be re-written \cite{invariants}:
\begin{equation}
A = \sum_i X^i \exp(-i\lambda_iL)
\label{amplitude}
\end{equation}
where the $\lambda_i$ are the eigenvalues of $H$
and the $X^i$ are hermitian (projection)
operators given by second-order polynomials (with real coefficients) of $H$:
\begin{equation}
X^i =
\frac{\prod_{j \neq i} (H-\lambda_j)}
                {\prod_{j \neq i} (\lambda_i-\lambda_j)}.
\label{lagrange}
\end{equation}
Clearly, the $X^i$ share the $\mu-\tau$ reflection symmetry of 
Eq.~(\ref{numass}). Comparison with the alternative approach 
in which the Hamiltonian is diagonalised before exponentiation allows 
us to identify the elements of the $X^i$ with the familiar
combinations of the lepton mixing matrix 
elements \cite{HPS33, invariants}:
\begin{equation}
X^i_{\alpha\beta} = U_{\alpha i}U^*_{\beta i}
\label{xi}
\end{equation}
(no summation over $i$ is implied).

The squared amplitude $|A_{\alpha\beta}|^2$ for $\alpha \neq \beta$
($\alpha=\beta$) gives the appearance (survival) probabilities 
as a function of neutrino energy and propagation distance 
and may be decomposed as a sum of two matrices:
\begin{equation}
{\cal P}_{\alpha\beta}(L/E)=|A_{\alpha\beta}|^2 
= {\cal S}_{\alpha\beta}(L/E)+{\cal A}_{\alpha\beta}(L/E)
\label{decomp}
\end{equation}
where the $CP$ (and $T$) symmetric and anti-symmetric terms are respectively:
\begin{eqnarray}
{\cal S}_{\alpha\beta}(L/E) &=& \delta_{\alpha\beta}
-4\sum_{i<j}K^{ij}_{\alpha\beta}\sin^2{(\Delta_{ij}L/2)} \cr
{\rm and}~{\cal A}_{\alpha\beta}(L/E) &=&
8J_{\alpha\beta}\sin{(\Delta_{12}L/2)}\sin{(\Delta_{23}L/2)}\sin{(\Delta_{31}L/2)}
\label{prob}
\end{eqnarray}
(the dependence on neutrino energy is implicit in the factors 
$\Delta_{ij}=(m^2_i-m^2_j)/2E$). The respective $CP$-even coefficients are:
\begin{equation}
K^{ij}_{\alpha\beta}
={\rm Re}(U_{\alpha i}U^*_{\beta i}U^*_{\alpha j}U_{\beta j})
={\rm Re}(X^i_{\alpha\beta}X^{j*}_{\alpha\beta}),
\label{Kdef}
\end{equation}
and the $CP$-odd one is:
\begin{equation}
J_{\alpha\beta}
={\rm Im}(U_{\alpha i}U^*_{\beta i}U^*_{\alpha j}U_{\beta j})
={\rm Im}(X^i_{\alpha\beta}X^{j*}_{\alpha\beta})
=\epsilon_{\alpha\beta}{\cal J}
\label{Jdef}
\end{equation}
where the flavour-antisymmetric matrix $\epsilon$ is
\begin{eqnarray}
\epsilon
= \left( \matrix{ 0  &  1 & -1 \cr
                 -1  &  0 &  1 \cr
     \hspace{2mm} 1 \hspace{2mm} &
          \hspace{2mm} -1 \hspace{2mm} &
               \hspace{2mm} 0  \hspace{2mm} \cr } \right)
\label{antisymm}
\end{eqnarray}
and ${\cal J}$ is Jarlskog's $CP$- and $T$-odd invariant. For anti-neutrinos, 
${\cal P}_{\overline{\alpha}\overline{\beta}}(L/E)$ is given by the same
formulation modified by ${\cal J}\rightarrow -{\cal J}$.

As already noted, the projection operators $X^i$
are symmetric under $\mu-\tau$ reflection:
\begin{equation}
X^i\rightarrow (X^i)'\equiv \Emt X^i\Emt^{\dag}=(X^i)^*
\label{symmProj}
\end{equation}
where $\Emt$ was defined in Eq.~(\ref{permutation}).
Hence, the real matrices $K^{ij}$ and $J$ derived from them 
are respectively symmetric (anti-symmetric) under $\mu-\tau$ 
flavour exchange (as well as under the $CP$-transformation), and are both
therefore symmetric under full $\mu-\tau$ reflection
\footnote{The complex matrices (in flavour-space) of plaquettes 
\cite{bjorken} $\Pi^{ij}=K^{ij}+iJ~(i<j)$ are symmetric under $\mu-\tau$ reflection,
themselves having the form Eq.~(\ref{numass}).}. Hence, the $CP$-even and 
$CP$-odd components of ${\cal P}(L/E)$ defined 
in Eq.~(\ref{prob}) are separately $\mu-\tau$ reflection symmetric,
${\cal S}(L/E)$ having the analogous form to that of 
Eq.~(\ref{numass}) with all parameters real, while 
${\cal A}(L/E)$, is proportional to $\epsilon{\cal J}$ whose factors flip sign
under $\mu-\tau$ flavour exchange and under $CP$ respectively.

By unitarity (applied separately for $\nu$ and $\overline{\nu}$), 
the rows and columns of ${\cal S}(L/E)$ each sum to unity (at each value 
of $L/E$), while those of ${\cal A}(L/E)$ each sum to zero. By $CPT$ 
symmetry, ${\cal S}(L/E)$ is a symmetric matrix 
(${\cal S}={\cal S}^T$; for brevity, we drop the dependence on $L/E$), 
while ${\cal A}(L/E)$ is anti-symmetric $({\cal A}=-{\cal A}^T)$. In the 
general vacuum case, these constraints reduce the number of independent 
functions of $L/E$ which are needed to specify ${\cal P}(L/E)$ to three 
$CP$-even (eg.~${\cal S}_{\alpha\beta}(L/E),\,\alpha<\beta$) and one
$CP$-odd (eg.~${\cal A}_{e\mu}(L/E)$). Hence survival probabilities 
can be written entirely 
in terms of the $CP$-conserving parts of appearance probabilities 
(${\cal P}_{\alpha\alpha}=1-{\cal S}_{\alpha\beta}-{\cal S}_{\alpha\gamma},\,
\alpha\neq\beta\neq\gamma$) and vice versa 
(${\cal S}_{\alpha\beta}=\frac{1}{2}(1-{\cal P}_{\alpha\alpha}
-{\cal P}_{\beta\beta}+{\cal P}_{\gamma\gamma})$).

Adding the constraints of $\mu-\tau$ reflection symmetry further reduces 
the number of independent $CP$-even functions of $L/E$ to two and we may 
therefore write:
\begin{eqnarray}
{\cal P}
= \left( \matrix{ 1-2{\cal S}_{e\mu}       &  {\cal S}_{e\mu} & {\cal S}_{e\mu} \cr
                  {\cal S}_{e\mu} &  1-{\cal S}_{e\mu}-{\cal S}_{\mu\tau} &  {\cal S}_{\mu\tau} \cr
     \hspace{2mm} {\cal S}_{e\mu} \hspace{2mm} &
          \hspace{2mm} {\cal S}_{\mu\tau} \hspace{2mm} &
               \hspace{2mm} 1-{\cal S}_{e\mu}-{\cal S}_{\mu\tau}  \hspace{2mm} \cr } \right)
+{\cal A}
\label{symm1}
\end{eqnarray}
where ${\cal A}=\epsilon{\cal A}_{e\mu}$ and $\epsilon$ was given in 
Eq.~(\ref{antisymm}), or equivalently, in terms of survival probabilities 
(${\cal S}_{\alpha\alpha}={\cal P}_{\alpha\alpha}$):
\begin{eqnarray}
{\cal P}
= \left( \matrix{ {\cal S}_{ee}  &  \frac{1}{2}(1-{\cal S}_{ee}) & \frac{1}{2}(1-{\cal S}_{ee}) \cr
                  \frac{1}{2}(1-{\cal S}_{ee})     &  {\cal S}_{\mu\mu} & 
		  \frac{1}{2}(1+{\cal S}_{ee}-2{\cal S}_{\mu\mu}) \cr
     \hspace{2mm} \frac{1}{2}(1-{\cal S}_{ee}) \hspace{2mm} &
          \hspace{2mm} \frac{1}{2}(1+{\cal S}_{ee}-2{\cal S}_{\mu\mu}) \hspace{2mm} &
               \hspace{2mm} {\cal S}_{\mu\mu}  \hspace{2mm} \cr } \right)
+{\cal A}\,.
\label{symm2}
\end{eqnarray}
Hence there are only three functions to evaluate in order to 
completely specify the matrix ${\cal P}(L/E)$.

For completeness, we provide here the expressions for the various
components of $\cal P$, (noting from above that ${\cal S}_{e\mu}$ and 
${\cal S}_{\mu\tau}$ are not independent of ${\cal S}_{ee}$ and 
${\cal S}_{\mu\mu}$):
\begin{eqnarray}
{\cal S}_{ee}(L/E)&=&1-4\sum_{i<j}u_i^2u_j^2\sin^2{(\Delta_{ij}L/2)}\label{pee1}\\
{\cal S}_{\mu\mu}(L/E)&=&1-\sum_{i<j}(1-u_i^2)(1-u_j^2)\sin^2{(\Delta_{ij}L/2)}\label{pee2}\\
{\cal S}_{e\mu}(L/E)&=&2\sum_{i<j}u_i^2u_j^2\sin^2{(\Delta_{ij}L/2)}\label{pee3}\\
{\cal S}_{\mu\tau}(L/E)&=&\sum_{i\neq j\neq k}(u_k^2-u_i^2u_j^2)\sin^2{(\Delta_{ij}L/2)}\label{pee4}\\
{\cal A}_{e\mu}(L/E)&=&4(u_1u_2u_3)\sin{(\Delta_{12}L/2)}\sin{(\Delta_{23}L/2)}\sin{(\Delta_{31}L/2)}\label{pee5}.
\end{eqnarray}
Of course, only two elements of $\underline{u}$ are independent, and 
the third may be simply determined from a given pair using the normalisation
of $\underline{u}$.

\vspace{7mm}
\ni \boldmath{\bf 5 Matter Effects}\unboldmath
\vspace{2mm}
\nl Matter effects are of considerable phenomenological importance. In this
section, we discuss how they affect the predictions of $\mu-\tau$ reflection
symmetry under quite general circumstances.
When (anti-)neutrinos propagate in matter, the Hamiltonian in the
flavour basis is modified: $H \rightarrow \tilde{H}$, with
\begin{equation}
\tilde{H}=H\pm{\rm diag}(a(L),0,0)
\label{matterHam}
\end{equation}
where $a(L)=\sqrt{2}G_F N_e(L)$, $N_e(L)$ is the (in general 
position-dependent) electron number density of the matter, and the
``$-$'' sign applies in the case of anti-neutrinos. 
This modifies both the eigenvalues and eigenstates of the Hamiltonian 
in a non-trivial way, in general as a function of $L/E$. However, 
in the case of a $\mu-\tau$ reflection symmetric Hamiltonian, 
Eq.~(\ref{numass}), its symmetry is unaffected, 
as only the parameter $z$ is changed ($z\rightarrow \tilde{z}=z\pm2Ea(L)$). 
Hence, {\em matter effects respect $\mu-\tau$ reflection symmetry} and
any result which is a consequence of the symmetry, is unaltered when 
neutrinos pass through matter. 

We first consider the most general case of arbitrary matter profile, and
no $\mu-\tau$ reflection symmetry. It is still convenient to 
decompose the matrix ${\cal P}(L/E)$ of survival and appearance probabilities
into $CP$-even and $CP$-odd parts, ${\cal S}(L/E)$ and ${\cal A}(L/E)$ 
respectively. Without solving the Schroedinger Equation for the particular 
$L$-dependent Hamiltonian of Eq.~(\ref{matterHam}) the functional dependences 
of ${\cal S}$ and ${\cal A}$ on $L/E$ are unknown. Moreover, in arbitrary 
matter, several of the 
familiar constraints of the vacuum case are lost, in particular
because $CPT$ symmetry does not in general hold. For example, the matrix 
${\cal S}(L/E)$ is no longer symmetric, and ${\cal A}(L/E)$ is no longer 
anti-symmetric; the survival probabilities ${\cal P}_{\mu\mu}$ 
and ${\cal P}_{\tau\tau}$ acquire $CP$-violating contributions \cite{minakata}.
However, we can still learn a considerable amount from general principles:
appealing to unitarity, as in the previous section, we see that in general, 
there are now four independent $CP$-even functions of $L/E$, eg.~we may write:
\begin{eqnarray}
{\cal S}
= \left( \matrix{ 1-{\cal S}_{e\mu}-{\cal S}_{e\tau}   &  {\cal S}_{e\mu} & {\cal S}_{e\tau} \cr
                  {\cal S}_{\mu e} &  1-{\cal S}_{\mu e}-{\cal S}_{\mu\tau} &  {\cal S}_{\mu\tau} \cr
     \hspace{2mm} {\cal S}_{e\tau}-({\cal S}_{\mu e}-{\cal S}_{e\mu}) \hspace{2mm} &
          \hspace{2mm} {\cal S}_{\mu\tau}+({\cal S}_{\mu e}-{\cal S}_{e\mu}) \hspace{2mm} &
               \hspace{2mm} 1-{\cal S}_{e\mu}-{\cal S}_{\mu\tau}  \hspace{2mm}\cr } \right).
\label{SarbMatt}
\end{eqnarray}
Similarly, there are three $CP$-odd ones (we use the fact that ${\cal A}_{ee}=0$
\cite{kuo}) eg.:
\begin{eqnarray}
{\cal A}
= \left( \matrix{ 0      &  {\cal A}_{e\mu}   & -{\cal A}_{e\mu} \cr
        {\cal A}_{\mu e} &  {\cal A}_{\mu\mu} &  -{\cal A}_{\mu e}-{\cal A}_{\mu\mu} \cr
 \hspace{2mm} -{\cal A}_{\mu e} \hspace{2mm} &
          \hspace{2mm} -{\cal A}_{e\mu}-{\cal A}_{\mu\mu} \hspace{2mm} &
               \hspace{2mm} {\cal A}_{e\mu}+{\cal A}_{\mu e}+{\cal A}_{\mu\mu} \hspace{2mm} \cr } \right).
\label{AarbMatt}
\end{eqnarray}
The $\mu-\tau$ reflection symmetry makes the following 
predictions\footnote{It is interesting to note that in the case of the element 
${\cal P}_{\mu\tau}={\cal S}_{\mu\tau}+{\cal A}_{\mu\tau}$, the $\mu-\tau$
exchange operator, $\Emt$, has the same effect as the time-reversal 
operator, $T$, and that the resulting constraints on these matrix elements
due to \mtrs\ are identical to those of $CPT$ symmetry. This accidental 
mimicking of $CPT$ symmetry is, however, restricted to the ${\cal P}_{\mu\tau}$ 
and ${\cal P}_{\tau\mu}$ elements, and \mtrs\ generally leads to constraints 
distinct from those of $CPT$ symmetry.}:
\begin{eqnarray}
{\cal S}_{e\mu}(L/E)={\cal S}_{e\tau}(L/E)&;&
{\cal S}_{\mu e}(L/E)={\cal S}_{\tau e}(L/E)\label{matterPred1}\\
{\cal S}_{\mu\tau}(L/E)={\cal S}_{\tau\mu}(L/E)&;&
{\cal S}_{\mu\mu}(L/E)={\cal S}_{\tau\tau}(L/E)\label{matterPred2}\\
{\cal A}_{e\mu}(L/E)=-{\cal A}_{e\tau}(L/E)&;&
{\cal A}_{\mu e}(L/E)=-{\cal A}_{\tau e}(L/E)\label{matterPred3}\\
{\cal A}_{\mu\tau}(L/E)=-{\cal A}_{\tau\mu}(L/E)&;&
{\cal A}_{\mu\mu}(L/E)=-{\cal A}_{\tau\tau}(L/E)\label{matterPred4}
\end{eqnarray}
(clearly, some though not all of them are redundant relative to the 
constraints of unitarity already invoked, eg.~Eq.~(\ref{matterPred3})).
Although following simply from \mtrs, we have verified these 
predictions in detail using the elegant formalism \cite{kimura} for 
arbitrary matter profiles.

It is easy to see that the constraints from \mtrs\ given in 
Eqs.~(\ref{matterPred1}) and (\ref{matterPred2}) reduce ${\cal S}(L/E)$ 
to the form already given in Eq.~(\ref{symm1}), so that now it has only two 
independent components, as before. Similarly, constraints (\ref{matterPred3}) 
and (\ref{matterPred4}) simplify the form of ${\cal A}(L/E)$ so that 
its complete specification now requires two functions to be determined, 
eg.~${\cal A}_{\mu\tau}(L/E)$ and ${\cal A}_{\mu\mu}(L/E)$
(or whichever are most convenient). Now, (using unitarity)
we may write:
\begin{eqnarray}
{\cal A}
= \left( \matrix{ 0      &  {\cal A}_{\mu\tau}-{\cal A}_{\mu\mu}   & {\cal A}_{\mu\mu}-{\cal A}_{\mu\tau} \cr
   -{\cal A}_{\mu\tau}-{\cal A}_{\mu\mu} &  {\cal A}_{\mu\mu} &  {\cal A}_{\mu\tau} \cr
 \hspace{2mm} {\cal A}_{\mu\tau}+{\cal A}_{\mu\mu} \hspace{2mm} &
          \hspace{2mm} -{\cal A}_{\mu\tau} \hspace{2mm} &
               \hspace{2mm} -{\cal A}_{\mu\mu} \hspace{2mm} \cr } \right).
\label{AarbMTRS}
\end{eqnarray}
For $T$-symmetric matter profiles, ${\cal A}_{\mu\mu}=0$ \cite{kimura}, 
and the symmetry of ${\cal A}(L/E)$ reduces to that of the vacuum case,
Eq.~(\ref{antisymm}).

\vspace{7mm}
\ni \boldmath{\bf 6 Application to Atmospheric Neutrinos}\unboldmath
\vspace{2mm}
\nl The observation from atmospheric neutrinos \cite{atmos} that 
$|U_{\mu3}|\simeq 1/\sqrt{2}$, was important to the argument in Section 1 that 
Nature respects strong $\mu-\tau$ universality, at least approximately. 
However, $|U_{\mu3}|= 1/\sqrt{2}$ is not a prediction of $\mu-\tau$ reflection 
symmetry in general. On the contrary, from Eq.~(\ref{normalisation}), 
if the symmetry is good, we have that 
$|U_{\mu3}|=\left(\frac{1-|U_{e3}|^2}{2}\right)^{1/2}$, which gives 
$|U_{\mu3}|=1/\sqrt{2}$ only if $U_{e3}=0$, 
a possibility which is not excluded, but which is by no means essential. 
In fact, if $0\le|U_{e3}|~\simlt~0.17$ as is currently thought to be the case
\cite{chooz, paloverde}, then $\mu-\tau$ reflection symmetry predicts
\begin{equation}
0.696~\simlt~|U_{\mu3}|=|U_{\tau3}|\leq 1/\sqrt{2},
\label{Umu3pred}
\end{equation}
a rather precise prediction, and certainly in agreement with the data.

One piece of experimental data which we have not used in the present paper,
is the non-observation of electron neutrino disappearance (or appearance)
in the atmospheric neutrino oscillation data \cite{superKElec}.
It is tempting to think that the closeness of the measured atmospheric 
electron neutrino rate to the expected value, independent of zenith angle, 
is evidence for the fact that $|U_{e3}|<<1$.
We will show here that under $\mu-\tau$ reflection symmetry, and in the 
(realistic) approximation 
of an initial flux ratio of muon neutrinos to electron neutrinos, 
$\phi_{\nu_{\mu}}/\phi_{\nu_{e}}=2$,
the measured electron neutrino survival probability is close to unity, 
{\em independent of the value of $|U_{e3}|$}, and hence that there is 
essentially no sensitivity to $|U_{e3}|$ in the rate of atmospheric electron 
neutrino events.

From Eq.~(\ref{symm1}), it can be seen that a prediction of 
$\mu-\tau$ reflection symmetry is that the $CP$-conserving part of the
appearance probability for electron neutrinos in a muon neutrino beam, 
${\cal S}_{\mu e}$, is exactly half of the disappearance probability 
from an electron neutrino beam, $1-{\cal P}_{ee}$, for all $L/E$ 
(it is even true for arbitrary matter profiles, Eq.~(\ref{matterPred1})).
In the limit $\phi_{\nu_{\mu}}/\phi_{\nu_{e}}=2$, these two terms 
compensate each other exactly in the measured rate of atmospheric electron 
neutrinos. Defining the cross-section-weighted neutrino flavour fractions
$r=\sigma_{\nu_{e}}\phi_{\nu_{e}}/(\sigma_{\nu_{e}}\phi_{\nu_{e}}+\sigma_{\overline{\nu}_{e}}\phi_{\overline{\nu}_{e}})$,
$s=\sigma_{\nu_{e}}\phi_{\nu_{\mu}}/(\sigma_{\nu_{e}}\phi_{\nu_{e}}+\sigma_{\overline{\nu}_{e}}\phi_{\overline{\nu}_{e}})$
and
$\overline{s}=\sigma_{\overline{\nu}_{e}}\phi_{\overline{\nu}_{\mu}}/(\sigma_{\nu_{e}}\phi_{\nu_{e}}+\sigma_{\overline{\nu}_{e}}\phi_{\overline{\nu}_{e}})$,
the ratio of the observed electron neutrino rate to that expected with
no oscillations is given by:
\begin{eqnarray}
R_e(L/E)&=&r[{\cal P}_{ee}(L/E)
         +(\phi_{\nu_{\mu}}/\phi_{\nu_{e}}){\cal P}_{\mu e}(L/E)]
         +(1-r)[\nu\leftrightarrow \overline{\nu}]\cr
        &=&r+(1-r)+s{\cal A}_{\mu e}+\overline{s}{\cal A}_{\overline{\mu} \overline{e}}
	\quad {\rm for}~\phi_{\nu_{\mu}}/\phi_{\nu_{e}}=\phi_{\overline{\nu}_{\mu}}/\phi_{\overline{\nu}_{e}}=2\cr
&\simeq&1
\label{peeAtmos}
\end{eqnarray}
where the $CP$-violating effects are suppressed by the smallness of 
$U_{e3}$ and $\Delta m^2_{12}$, which would explain why present data
are consistent with unity. While such compensation theorems have been 
noted previously by several authors for specific models 
\cite{HPS33, ahluwalia},
we emphasise here the perfect compensation of the $CP$-even terms 
for the generality of models with $\mu-\tau$ reflection symmetry, and
for arbitrary matter profiles. We note furthermore, that although naively 
there is a partial cancellation in Eq.~(\ref{peeAtmos}) between the terms 
involving ${\cal A}_{\mu e}$ and ${\cal A}_{\overline{\mu} \overline{e}}$ 
(owing to a relative minus sign in their dependences on ${\cal J}$), 
matter effects have different influences on their respective $L/E$ 
dependences, and the cancellation may be incomplete, even if $s=\overline{s}$.

Any observed deviation from $R_e(L/E)=1$ is therefore evidence for one of 
three things: i) a deviation from $(\phi_{\nu_{\mu}}/\phi_{\nu_{e}})=2$ 
   and/or from $(\phi_{\overline{\nu}_{\mu}}/\phi_{\overline{\nu}_{e}})=2$, 
   either of which would appear in the leading oscillations, given sufficient 
   $L/E$ resolution; 
ii) a deviation from $\mu-\tau$ reflection symmetry, which would also appear
in the leading oscillation;
iii) $CP$ violation, which is potentially distinguishable from i) and ii)
as it would be manifest only on the longer distance scale of 
$\Delta m^2_{12} L/4E{\not \ll} 1$.

\vspace{7mm}
\ni \boldmath{\bf 7 Conclusions}\unboldmath
\vspace{2mm}
\nl Current data from atmospheric, solar and reactor neutrino oscillation 
experiments are all consistent with the hypothesis of $\mu-\tau$ reflection 
symmetry for the leptonic sector of the Standard Model (augmented with
neutrino masses and standard MNS mixings). 
The $\mu-\tau$ reflection transformation is the product
of the $\mu-\tau$ exchange operator, $E_{\mu\tau}$, applied to the neutrino
flavour eigenstates, and the $CP$ 
operator. $CP\Emt$ is clearly a unitary transformation and is
also Hermitian with $(CP\Emt)^2=I$, so that its eigenvalues are $\pm 1$.
If \mtrs\ is a good symmetry of the neutrino oscillation Hamiltonian, 
one would expect the $CP\Emt$ quantum number to be conserved in such
processes. One might call this quantum number ``$\mu-\tau$ reflectivity''
or simply ``{\em mutativity}''. The eigenstates of mutativity are clearly:
\begin{eqnarray}
{\rm mutativity} +1&:& \frac{1}{\sqrt{2}}(\numu+\nutaubar)~{\rm and}~
\frac{1}{\sqrt{2}}(\numubar+\nutau)\\
{\rm mutativity} -1&:& \frac{1}{\sqrt{2}}(\numu-\nutaubar)~{\rm and}~
\frac{1}{\sqrt{2}}(\numubar-\nutau)
\end{eqnarray}
although it is not easy to see how one might prepare a beam of neutrinos 
in such an eigenstate to test directly the conservation law in a non-trivial
way.

As we have explored in the main body of this paper, in addition to the above 
postulated conservation law, the hypothesis of \mtrs\ makes a set of 
specific predictions about the $CP$-even and $CP$-odd parts of neutrino 
appearance and survival probabilities. Further detailed measurements
are required to falsify this hypothesis, by finding deviations from
the predictions, Eqs.~(\ref{matterPred1})-(\ref{matterPred4}), 
valid even in the presence of terrestrial matter, and 
Eqs.~(\ref{equality}), (\ref{normalisation}), 
(\ref{orthogonality2}), (\ref{JCP}), (\ref{Umu3pred}) and (\ref{peeAtmos}).

\vspace{7mm}
\ni \boldmath{\bf Acknowledgements}\unboldmath
\vspace{2mm}
\nl It is a pleasure to thank Tom Weiler for useful suggestions. This work
was supported by the UK Particle Physics and Astronomy Research Council
(PPARC).

\newpage
\ni \boldmath{\bf Appendix A: $\mu-\tau$ Reflection Symmetry in the Neutrino Mass Basis}\unboldmath
\vspace{2mm}
\nl It is also interesting to consider the manifestations of \mtrs\ in
another special weak basis, namely that in which the neutrino mass matrix 
is diagonal (the neutrino mass basis). In this basis, the lepton mixing
degrees of freedom are all contained in the charged lepton mass matrix-squared 
(the weak interaction, and the neutrino mass matrix are both diagonal
by definition in this basis), which is then given by
\begin{equation}
M^2_{\ell}= U^{\dag}D^2_{\ell}U
\label{leptondef}
\end{equation}
where $D^2_{\ell}={\rm diag}(m^2_e,m^2_{\mu},m^2_{\tau})$, from which we obtain
\begin{equation}
(M^2_{\ell})_{ij}=\left(\frac{m^2_{\mu}+m^2_{\tau}}{2}\right)\delta_{ij}+
\left[m^2_e-\left(\frac{m^2_{\mu}+m^2_{\tau}}{2}\right)\right]u_iu_j+
i\epsilon_{ijk}\left(\frac{m^2_{\tau}-m^2_{\mu}}{2}\right)u_k,
\label{leptonmass}
\end{equation}
whose real part is symmetric under the interchange $\mu\leftrightarrow\tau$ 
and whose imaginary part is anti-symmetric (in this basis, the rows and 
columns of $M^2_{\ell}$ are labelled by the neutrino mass eigenstate indices, 
the flavour indices labelling only the charged lepton mass eigenvalues). 
This matrix explicitly respects $\mu-\tau$ reflection symmetry, as the effects 
of $\mu\leftrightarrow\tau$ flavour exchange are clearly reversed by a $CP$ 
transformation, which is again equivalent to complex conjugation. 
The fact that the symmetry is manifest also in this basis demonstrates
that the symmetry is not simply an accident in the flavour basis 
considered in the main text.
$M^2_{\ell}$ encodes 5 degrees of freedom -  the three charged lepton 
masses, and any pair of elements of the $\nu_e$ flavour eigenstate, 
$\underline{u}$.

\end{document}